# Estimation of the Perturbing Accelerations Induced on the LARES Satellite by Neutral Atmosphere Drag

C. Pardini[a*], L. Anselmo[a], D.M. Lucchesi[a,b,c] and R. Peron[b,c]

[a] Space Flight Dynamics Laboratory, Istituto di Scienza e Tecnologie dell'Informazione (ISTI), Consiglio Nazionale delle Ricerche (CNR) – Pisa, Italy
[b] Experimental Gravitation Section, Istituto di Astrofisica e Planetologia Spaziali (IAPS), Istituto Nazionale di Astrofisica (INAF) – Roma, Italy
[c] Istituto Nazionale di Fisica Nucleare (INFN), Sezione di Roma Tor Vergata, Via della Ricerca Scientifica 1 – Roma, Italy

[*] *Corresponding author. Tel.*: +39-050-315-2987; *Fax*: +39-050-315-2040;
  *Postal address*: ISTI/CNR, Via G. Moruzzi 1, 56124, Pisa, Italy;
  *E-mail*: carmen.pardini@isti.cnr.it

**Abstract**

The laser-ranged satellite LARES is expected to provide new refined measurements of relativistic physics, as well as significant contributions to space geodesy and geophysics. The very low area-to-mass ratio of this passive and dense satellite was chosen to reduce as much as possible the disturbing effects of non-gravitational perturbations. However, because of its height, about 1450 km compared with about 5800-5900 km for the two LAGEOS satellites, LARES is exposed to a much stronger drag due to neutral atmosphere.

From a precise orbit determination, analyzing the laser ranging normal points of LARES over a time span of about 3.7 years with the GEODYN II (NASA/GSFC) code, it was found an average semi-major axis decay rate of $-0.999$ m per year, corresponding to a non-conservative net force acting nearly opposite to the velocity vector of the satellite and with a mean along-track acceleration of $-1.444 \times 10^{-11}$ m/s$^2$.

By means of a modified version of the SATRAP (ISTI/CNR) code, the neutral drag perturbation acting on LARES was evaluated over the same time span, taking into account the real evolution of solar and geomagnetic activities, with five thermospheric density models (JR-71, MSIS-86, MSISE-90, NRLMSISE-00 and GOST-2004). All of them were able to model most of the observed semi-major axis decay, with differences among the average drag coefficients smaller than 10%. Moreover, when the same models (JR-71 and MSIS-86) were used within GEODYN in a least-squares fit of the tracking data, the differences between the average drag coefficients estimated with SATRAP and GEODYN were of the order of 1% or less. Adding to this a further independent check carried out analyzing the orbital decay of a passive spherical satellite (Ajisai) just 40 km higher than LARES, it was then concluded that some of the currently best models developed for neutral atmosphere, within their uncertainties and range of applicability, were able to account for most ($\approx 98.6\%$) of the observed semi-major axis decay of LARES.

Finally, after modeling the neutral atmosphere drag in GEODYN, a residual semi-major axis decay, corresponding to an average along-track acceleration of about $-2 \times 10^{-13}$ m/s$^2$ (i.e. $\approx 1/72$ of neutral drag), was detected as well. It is probably linked to thermal thrust effects (Nguyen and Matzner, 2015; Brooks and Matzner, 2016), but further and more detailed investigations, including the detection of the signature of the periodic terms, will be needed in order to characterize such smaller non-gravitational perturbations.

## 1. Introduction

LARES (LAser RElativity Satellite) was launched from Kourou, in French Guiana, on 13 February 2012, with the qualification flight of the VEGA launcher. With a radius of 18.2 cm and a mass of 386.80 kg, this completely passive sphere, made of tungsten alloy and uniformly hosting on its surface 92 corner cube laser retro-reflectors, is the densest object in orbit in the Solar System, with an area-to-mass ratio $A/M = 2.69 \times 10^{-4}$ m$^2$/kg (Paolozzi and Ciufolini, 2013). Placed into a nearly circular orbit at a mean geodetic altitude of about 1454 km and with an inclination of 69.5°, LARES was conceived with the goal of measuring the Lense-Thirring effect predicted by General Relativity with accuracy of the order of 1% (Ciufolini et al., 2009). It also represents an essential component of the LAser RAnged Satellites Experiment (LARASE), whose main goal is to provide accurate measurements for the gravitational interaction in the weak-field and slow-motion limit by means of the laser tracking of satellites orbiting around the Earth (Lucchesi et al., 2015a).

The ambitious scientific objectives of the mission will be met only if, together with high quality laser tracking of the satellite, it will be possible to improve the dynamical models describing its motion, and that of other laser ranged satellites, concerning in particular the subtle effects of non-gravitational perturbations (Lucchesi et al., 2015b). Among these, the effects of the neutral atmosphere drag deserve a special attention in the case of LARES, the satellite having been placed in a relatively low orbit, where the residual Earth atmosphere cannot be ignored, compared with other perturbations whose modeling is relevant for meeting the scientific goals of the mission.

The reliable and detailed estimation of the perturbing accelerations induced on the LARES satellite by the neutral atmosphere drag is therefore the main purpose of this paper. The results obtained, when integrated in the LARASE precise orbit determination process, shall significantly contribute to the improvement of the dynamical model, reducing the unmodeled residuals and offering at the same time new insights on other non-gravitational perturbations, e.g. charged particle and thermal drag.

## 2. Neutral atmosphere drag

Atmospheric drag is the largest non-gravitational perturbation acting on satellites below the altitude of ~1000 km (Montenbruck and Gill, 2001), and also at the LARES altitude must be taken into account to explain the details of the orbital evolution inferred from the extremely accurate tracking made possible by laser ranging. The most revealing feature of the drag force, which for a spherical body with uniform surface properties is directed opposite to the satellite's velocity relative to the atmosphere, is the dissipation of orbital energy, leading to a gradual decrease of the satellite's semi-major axis.

It is usual to express the drag force ($F_D$) in the form:

$$F_D = M\, a_D = -1/2\, \rho\, C_D A\, V_r^2\, V_r/V_r\,, \qquad (1)$$

in which $a_D$ is the corresponding acceleration, $M$ is the mass of the satellite, $\rho$ is the local atmospheric density, $V_r$ is the velocity of the satellite relative to the atmosphere, $A$ is the cross-sectional area of the satellite facing the airstream, and $C_D$ is the so-called drag coefficient. The values of the area and the mass are well known in the case of LARES, but the relative velocity, which depends on the complex dynamics of the Earth's atmosphere, may be affected by uncertainties larger than 1%. However, a reasonable approximation of $V_r$ is obtained with the assumption that the atmosphere co-rotates with the Earth, i.e.:

$$V_r = V - \omega_\oplus \times r ,  \qquad (2)$$

where $V$ is the satellite inertial velocity, $r$ is its position vector, and $\omega_\oplus$ is the Earth's angular velocity vector. In most situations this assumption leads to negligible errors, but during the relatively short duration (from several hours to a few days) of major geomagnetic storms, occurring approximately 5% of the time, the atmospheric winds may grow remarkably (Moe et al., 2004).

In general, however, the drag coefficient and the atmospheric density represent the principal causes of uncertainty in computing the drag force. This is the consequence of the fact that when $F_D$ or $a_D$ are measured, for instance through the observed orbital decay of a satellite, it is the product $\rho \cdot C_D$ which is determined, so any uncertainty in $C_D$ brings forth an uncertainty in the air density $\rho$, and vice versa. Therefore, for this and other reasons (Pardini et al., 2012), any semi-empirical atmospheric density model may be affected by significant errors. Nevertheless, if Eqs. (1) and (2) rightly fit the behavior of the drag force and the air density $\rho$ is appropriately described by a sufficiently sophisticated model, i.e. able to follow the changes induced by the varying solar and geomagnetic activity and by seasonal effects, a large fraction of the uncertainty in $\rho$, as well as other possible and smaller drag-like contributions from other perturbations, e.g. the charged particle and thermal drag in the case of LARES, might be incorporated in the value of $C_D$, if set as solve for parameter in the precise orbit determination process. In other words, the value of $C_D$ so determined would not necessarily correspond to the physical aerodynamic drag coefficient of the satellite, having to absorb the average density bias of the atmospheric model, during the considered integration arc, and other possible smaller drag-like influences (limited to the constant component, if $C_D$ were assumed constant as well over the integration arc). However, such a procedure would lead in any case to a better orbit determination and to smaller residuals, benefitting the primary goal of the mission, i.e. the General Relativity measurements.

Of course, if the aim of the measurements were the precise and accurate determination of air density as a function of the season and space weather conditions, an independent estimation of the physical drag coefficient would be needed, to be compared with the $C_D$ found as solve for parameter in the precise orbit determination process (Pardini et al., 2012). Nevertheless, as explained above, this is not presently necessary for the primary goals of the LARASE experiment.

## 3. The observed orbital decay of LARES

In order to reduce as much as possible the effects of the non-gravitational perturbations, LARES was designed and built with an area-to-mass ratio 2.6 times smaller than that of the twin LAser GEOdynamics Satellites (LAGEOS and LAGEOS II), placed at mean geodetic altitudes, respectively, of 5897 km and 5785 km, in 1976 and in 1992 (Lucchesi et al., 2015a; Visco and Lucchesi, 2016). This led, in fact, to reduced direct solar and Earth's albedo radiation pressures, but, due to the significantly lower altitude of LARES, neutral atmosphere drag was expected to be approximately 50 times stronger than for the two LAGEOS (Lucchesi et al., 2015a). The accurate modeling of such non-gravitational perturbation is therefore of paramount importance for LARES if the ambitious scientific goals of the mission have to be met in a credible way.

Analyzing the first 1.5 years of laser ranging data, Sośnica et al. (2013) measured an average decay of the LARES semi-major axis by −0.775 m per year. In a totally independent analysis, in terms of data set and time interval, we used the two-line elements sets (Hoots and Roehrich, 1980; Vallado and Crawford, 2008) determined by the US Strategic Command over 3.7 years, from 18 February 2012 to 22 October 2015, revealing an average secular decrease of the semi-major axis by

−0.952 m per year. Using the NASA/GSFC software package GEODYN II (Putney et al., 1990; Pavlis et al., 1998), we also carried out a precise orbit determination, based on the laser ranging data of LARES (normal points) provided by the International Laser Ranging Service (ILRS) (Pearlman et al., 2002). The analysis did not include in the dynamical model neither the neutral and charged atmosphere drag, nor the thermal effects, fitting the observables over 7-day orbit arcs, from 6 April 2012 to 25 December 2015. In this way, the detailed secular decrease of the LARES semi-major axis was uncovered, over more than 3.7 years (Figure 1). The corresponding average decay rate was −0.999 m per year (Lucchesi et al., 2016). Therefore, during the first 3.7 years in orbit, LARES was subjected to an average semi-major axis decay of 2.7 mm per day, and these results indicate that a non-conservative net force was acting on the satellite, with a mean along-track acceleration component of $-1.444 \times 10^{-11}$ m/s$^2$.

It is worth remembering that in the case of LAGEOS, and later on of LAGEOS II, the detection of a small but unexpected (and originally unexplained) semi-major axis decay (on average, −0.203 m per year and −0.239 m per year, respectively, in the 1994-2011 period (Sośnica, 2014)) lead to a significant physical investigation and modeling effort. Consequently, it was quite logical to assess the similarities and differences among the LARES and the LAGEOS satellites semi-major axis decay, as a first step to characterize, quantify and properly model the non-gravitational perturbations acting on the former.

It was known since the beginning of the 1980's (Afonso et al., 1980; Rubincam, 1980; 1982) that neutral atmosphere drag was not able to account for the LAGEOS orbital decay, being one order of magnitude too small. However, the considerably lower altitude of LARES put the attention on neutral atmosphere drag as the leading cause of the observed secular decrease of the semi-major axis, as even simple preliminary computations are able to show. In fact, assuming a circular orbit (the LARES eccentricity is less than $10^{-3}$) and that all the observed semi-major decay is due to neutral atmosphere drag, the average neutral atmospheric density ρ at the satellite altitude can be estimated as follows:

$$\rho = -\Delta a/[2\pi(C_D \cdot A/M)a^2] , \qquad (3)$$

where $a$ is the semi-major axis and $\Delta a$ represents its variation over one revolution (Larson and Wertz, 1992):

$$\Delta a = -2\pi(C_D \cdot A/M)\rho a^2 . \qquad (4)$$

Taking into account that the LARES nodal orbital period is about 6885 s and that during 1300 days the semi-major axis decreased by about −3.5 m (Figure 1), the average $\Delta a$ resulted to be $-2.15 \times 10^{-4}$ m per revolution. Substituting these values of $\Delta a$ in Eq. (3), together with $a = 7820.2$ km and $C_D = 3.5$ (a reasonable value for a sphere at such an altitude using the Jacchia 1971 atmospheric density model; see the following discussion regarding the Ajisai satellite), it was then possible to estimate the mean neutral atmosphere density, at the LARES altitude, needed to explain the observed semi-major axis decay. The value obtained was $\rho = 5.9 \times 10^{-16}$ kg/m$^3$.

For LARES the estimated density was not one order of magnitude greater than what should have been expected, as happened for the LAGEOS satellites, but was instead in sound agreement with the predictions of the thermospheric density models. For instance, assuming the Jacchia 1971 model (Zarrouati, 1987), at the altitude of LARES it predicts an average neutral density $\langle\rho\rangle = 3.48 \times 10^{-16}$ kg/m$^3$ with an exospheric temperature $T_\infty = 500$ K, $\langle\rho\rangle = 6.32 \times 10^{-16}$ kg/m$^3$ with $T_\infty = 1000$ K, and $\langle\rho\rangle = 1.20 \times 10^{-14}$ kg/m$^3$ with $T_\infty = 2000$ K. Therefore, even a very rough and preliminary estimation seems to indicate that, considering the appropriate environmental conditions (i.e. solar and geomagnetic activities) prevailing during the time interval considered, characterized by $\langle T_\infty \rangle = 971$ K, neutral atmosphere drag might explain most of the observed orbital decay.

## 4. Detailed modeling of the neutral atmosphere drag

In view of the central role played by neutral atmosphere drag among the non-gravitational forces acting on LARES, a detailed modeling of such perturbation was deemed necessary for a better characterization and estimation of the other non-gravitational forces, and for the success of the LARASE experiment. To do so, a modified version of the SATRAP tool, developed at ISTI/CNR (Pardini and Anselmo, 1994; Pardini et al., 2012), was used to model the neutral drag acceleration acting on LARES, as a function of time, taking into account the real evolution of solar and geomagnetic activities and the observed mean secular along-track acceleration. In fact, many atmospheric density models have been implemented in SATRAP over more than 20 years of development, including Jacchia-Roberts 1971 (JR-71; Cappellari et al., 1976), the Mass Spectrometer and Incoherent Scatter radar 1986 model (MSIS-86; Hedin, 1987), the Mass Spectrometer and Incoherent Scatter radar Extended 1990 (MSISE-90; Hedin, 1991), NRLMSISE-00 (Picone et al., 2002a; 2002b), developed at the US Naval Research Laboratory (NRL), and GOST-2004 (Volkov, 2004), issued by the State Committee on Standardization and Metrology of the Russian Federation.

All the models listed above, i.e. JR-71, MSIS-86, MSISE-90, NRLMSISE-00 and GOST-2004, were used within SATRAP to compute the components of the neutral drag acceleration on LARES in the reference system RSW, having the origin in the center of mass of the satellite and with three orthogonal axes aligned, respectively, along the radial direction, from the center of the Earth to the satellite (**R**), normal to the orbit plane, in the direction of the orbital angular momentum (**W**), and in the transverse (along-track) direction (**S**), lying on the orbital plane 90 degrees from **R**, i.e. nearly aligned with the satellite velocity vector. This analysis covered the first 3.7 years, from 6 April 2012 to 25 December 2015, and the drag coefficients $C_D$ were adjusted in order to reproduce, with each atmospheric density model, the observed average transverse acceleration component $\langle S \rangle = -1.444 \times 10^{-11}$ m/s$^2$. For the applicable time interval, Figure 2 shows the geodetic altitude, the exospheric temperature and the overall atmospheric density, while Figure 3 shows the concentration of the three main atomic species, i.e. helium (He), hydrogen (H) and oxygen (O). (Both figures were obtained with NRLMSISE-00.)

The results obtained are summarized in Figures 4-10 and in Table 1. In order to fit the mean secular along-track non-gravitational acceleration observed on LARES, the various thermospheric density models converged, as expected, to different average drag coefficients, from $\langle C_D \rangle = 3.71$ in the case of MSIS-86 to $\langle C_D \rangle = 4.21$ in the case of GOST-2004 (Table 1). The average among the five models was $\langle C_D \rangle = 3.88$, with a maximum discrepancy of 8.6%, while considering that NRLMSISE-00 was a refinement of MSIS-86 and MSISE-90, sharing with them a common heritage, the average among the three more independent thermospheric models JR-71, NRLMSISE-00 and GOST-2004 was $\langle C_D \rangle = 3.98$, with a maximum discrepancy of 5.8%. The biases among the models, up to 9% (6%) with respect to the averaged values and up to 13% (11%) among the models themselves, are not surprising, because they are fully consistent with their known uncertainties (Volkov et al., 2008; Yurasov et al., 2004; CIRA-2012 International Working Group, 2012). This fact should rather alert anyone involved in the modeling of the LARES non-gravitational perturbations that any along-track secular acceleration with a magnitude of the order of $2 \times 10^{-12}$ m/s$^2$, or less, i.e. ≤ 15% of the neutral drag acceleration, might be easily hidden in the uncertainties and biases of the neutral thermosphere models, and eventually absorbed in the estimation of the $C_D$ solve for parameter during the precise orbit determination process. Therefore, only a detailed and careful analysis of the radial R and out-of-plane W perturbation components, coupled with the short and long-term variations of the transverse component S, would be possibly able to identify the characteristic signatures of smaller non-gravitational perturbations distinct from neutral atmosphere drag.

Figure 4 plots the transverse drag component S found with JR-71, NRLMSISE-00 and GOST-2004, together with the eclipse periods, when part of the LARES orbit lies in the Earth's shadow, the daily and averaged solar flux at 10.7 cm, and the geomagnetic activity index $K_p$. The general correlation between the magnitude of the acceleration and the level of solar activity is quite apparent. Figure 5 shows instead the results obtained with all five models together. As more evident in Figures 7, 8 and 9, MSIS-86, MSISE-90 and NRLMSISE-00 shared the same signature, with a very good agreement in terms of S magnitude and time evolution, with lower peaks exceeding $-4 \times 10^{-11}$ m/s$^2$. JR-71 (Figure 6) still displayed a relatively good agreement with the S signature of the other three models, but with deeper lower peaks, sometime exceeding $-6 \times 10^{-11}$ m/s$^2$. The Russian model GOST-2004, however, behaved differently in response to the varying environmental conditions (Figure 10), presenting a quite distinct signature and smaller magnitude excursions of S, with an absolute value nearly always smaller than $3 \times 10^{-11}$ m/s$^2$ (i.e. $-3.176 \times 10^{-11} < S < -2.058 \times 10^{-12}$ m/s$^2$). Then, the time-varying part of the transverse acceleration S was characterized by different frequencies and amplitudes among the models, in particular between GOST-2004 and the American ones, being this a further relevant aspect to be considered when looking for the signature of other smaller non-gravitational perturbations in the residuals of the precise orbit determination process.

The other components of the neutral drag acceleration are also plotted in Figure 6 for JR-71, Figure 7 for MSIS-86, Figure 8 for MSISE-90, Figure 9 for NRLMSISE-00 and Figure 10 for GOST-2004. First of all it should be mentioned that the radial R and normal W components resulted to have, as expected, mean values close to zero, together with short and long-term oscillations quite smaller than the transverse component S. Again, there was a very good agreement among the time evolution signatures of MSIS-86, MSISE-90 and NRLMSISE-00, and, to a slightly lesser extent, of JR-71, with the radial component R mostly varying between $\pm 4.5 \times 10^{-14}$ m/s$^2$ and the normal component W mostly varying between $\pm 2.6 \times 10^{-12}$ m/s$^2$. GOST-2004 presented similar magnitude excursions, just a little bit smaller in the extreme values, but, as in case of the S component, the time evolution signature was distinctly different with respect to the other models considered.

Being more specific, the low frequency behavior, here corresponding to periods greater than 500 days, was dominated by the external inputs to the models themselves, i.e. the solar and geomagnetic activity evolution. However, the high frequency behavior, here corresponding to periods smaller than 500 days, was also much affected by the intrinsic way the models were built, to respond to the varying solar and geomagnetic activity, taking into account the changing position of the Sun due to the yearly Earth's revolution ($\dot{\lambda}$), the precession of the satellite orbital plane ($\dot{\Omega}$) and the precession of the satellite perigee ($\dot{\omega}$).

Concerning the radial drag component R, the American models (JR-71, MSIS-86, MSISE-90 and NRLMSISE-00), sharing a common development heritage, displayed signal periodicities of comparable amplitude for $\dot{\lambda} - \dot{\omega}$ (186.6 days), $\dot{\Omega} - \dot{\lambda}$ (133.5 days) and $\dot{\Omega} - \dot{\lambda} + \dot{\omega}$ (98.9 days), of approximately halved amplitude for $\dot{\Omega} - \dot{\lambda} - \dot{\omega}$ (205.4 days), and reduced by a factor 5÷6 (MSIS-86, MSISE-90 and NRLMSISE-00) or 8÷9 (JR-71) for $\dot{\Omega} + \dot{\lambda}$ (496.8 days). The Russian model (GOST-2004), on the other hand, displayed only one significant signal periodicity, for $\dot{\lambda} - \dot{\omega}$ (186.6 days), characterized by an amplitude nearly 4 times smaller than the corresponding spectral line of the American models.

Regarding the transverse drag component S, the three similar models MSIS-86, MSISE-90 and NRLMSISE-00 were dominated by a signal periodicity for $2\dot{\lambda}$ (182.6 days, i.e. six months), followed by a spectral line at $\dot{\lambda}$ (365.3 days, i.e. one year), with an amplitude reduced by nearly a factor 3, and by another line at $2(\dot{\Omega} - \dot{\lambda})$ (66.8 days), with an amplitude reduced by nearly a factor 6. The same periodicities, with an identical amplitude sequence, were found with JR-71, but the $2\dot{\lambda}$ line was smaller by 21%, the $\dot{\lambda}$ line was greater by 50%, and the $2(\dot{\Omega} - \dot{\lambda})$ line was greater by 13%. Even the high frequency signal obtained with GOST-2004 was dominated by the $2\dot{\lambda}$ and $\dot{\lambda}$ periodicities, but no other significant line characterizing the relative orientation of the orbital plane with respect to the Sun, like $2(\dot{\Omega} - \dot{\lambda})$, was present. Compared with NRLMSISE-00, the $2\dot{\lambda}$ line found with GOST-2004 was smaller by 17%, while the $\dot{\lambda}$ line was greater by 72%.

The simpler high frequency spectrum of GOST-2004 was particularly apparent in the case of the out-of-plane drag component W, were only one significant signal periodicity for $\dot{\omega}$ (381.7 days) was found. The same spectral line, but with double amplitude, was also present in the outputs of the American models, but four other lines of progressively greater amplitude were present as well at $\dot{\Omega} - 2\dot{\lambda}$ (97.8 days), $\dot{\Omega} - 2\dot{\lambda} + \dot{\omega}$ (77.8 days), $\dot{\Omega} + \dot{\lambda}$ (496.8 days) and $\dot{\Omega} - \dot{\lambda}$ (133.5 days). The latter was by far the dominant one, with amplitude from 6 (NRLMSISE-00) to 9 (JR-71) times greater than the following at $\dot{\Omega} + \dot{\lambda}$, being the remaining three lines still smaller and comparable among them.

In conclusion, it is quite apparent the need to carefully analyze the RSW signature of neutral drag, according to various atmosphere density models, when looking for other smaller non-gravitational perturbations acting on LARES.

## 5. An independent check of neutral drag at the LARES altitude

A completely independent check of neutral atmosphere drag at the LARES altitude was possible thanks to another passive spherical satellite launched by Japan for geodetic research on 12 August 1986: Ajisai. With a diameter of 215 cm and a mass of 685 kg, the hollow sphere has an area-to-mass ratio $A/M = 5.30 \times 10^{-3}$ m$^2$/kg, i.e. 19.70 times that of LARES, being therefore much more sensitive to non-gravitational perturbations, like atmospheric drag and radiative forces. The surface of the satellite is basically silica ($SiO_2$) in composition, being completely covered with 318 mirrors for reflecting sunlight and 1436 quartz corner cube retro-reflectors for reflecting laser beams (Sengoku et al., 1995). The nearly circular orbit has a semi-major axis of 7866.5 km, a mean geodetic altitude of 1494 km and an inclination of 50.0 deg. Ajisai is therefore 40 km higher than LARES.

The secular orbital decay of Ajisai due to neutral atmosphere drag is known since its launch and has been used over the years to check the predictions of several atmospheric models under various conditions of solar and geomagnetic activity (Pardini and Anselmo, 2001; Pardini et al., 2006). During the approximately 3.7 years, from 6 April 2012 to 25 December 2015, discussed for LARES in the previous section, the average secular decrease of the semi-major axis of Ajisai was −14.041 m per year, as obtained with the two-line elements sets issued by the US Strategic Command. Therefore, the magnitude of the observed secular semi-major axis decay was about 14.055 times larger than for LARES. For the applicable time interval, Figure 11 shows the geodetic altitude, the exospheric temperature and the overall atmospheric density, while Figure 12 shows the concentration of the three main atomic species, i.e. helium (He), hydrogen (H) and oxygen (O). As in the LARES case, both figures were obtained with NRLMSISE-00.

In order to reproduce the observed mean secular decrease of the semi-major axis, the needed average transverse component $\langle S \rangle$ was found to be $-2.013 \times 10^{-10}$ m/s$^2$. Figure 13 shows the corresponding neutral drag components computed with JR-71, obtained with $\langle C_D \rangle = 3.41$, Figure 14 shows the output of NRLMSISE-00, obtained with $\langle C_D \rangle = 3.20$, and Figure 15 presents the results of the GOST-2004 computations, obtained with $\langle C_D \rangle = 3.34$ (the results of MSIS-86 and MSISE-90 are not shown again, being practically coincident with those of NRLMSISE-00). Taking the average between JR-71, NRLMSISE-00 and GOST-2004, the mean drag coefficient resulted to be 3.32, with a maximum discrepancy of 3.5%. Again, as discussed in the case of LARES, these differences are well below the acknowledged uncertainties of the models, so neutral atmosphere drag alone is able to account for the observed secular semi-major axis decay of Ajisai, according to the current knowledge of the circumterrestrial environment.

The drag coefficients compatible with the observed secular decay of the Ajisai semi-major axis, in the considered time interval, were also very close to the "theoretical" value (3.16) estimated by following the approach outlined in Afonso et al. (1985), taking into account a hyperthermal free molecular flow, a diffuse reemission process of the incident particles, the neutral atmosphere atomic species composition, their mean thermal velocities, the satellite relative velocity with respect to the atmosphere and the chemical composition of the satellite surface. Therefore, the thermospheric models were consistent among them and probably were not affected by significant density biases (in particular NRLMSISE-00) during the time interval and for the orbit considered. It should be however pointed out that repeating the estimation of the "theoretical" drag coefficient for LARES produced nearly the same value (3.18) obtained for Ajisai. The secular semi-major axis decay analysis, instead, obtained considerably higher $\langle C_D \rangle$ values for LARES compared with Ajisai, and this behavior was consistently shared by all the models, with JR-71 presenting an increase by 15.8%, NRLMSISE-00 by 18.1% and GOST-2004 by 26.0%.

Independently from the fact of considering realistic or not the above mentioned "theoretical" drag coefficient estimates, the average difference of approximately 20% between the "determined" drag coefficients of LARES and Ajisai, both passive spheres at similar altitudes, needed to be addressed. The following explanations were considered:

1. The effect of some significant and unmodeled radiative force able to induce a secular change, averaged over nearly four years, of the semi-major axis;
2. The effect of charged particle drag in contributing to the semi-major axis secular decay;
3. Some specific properties of LARES leading to a higher than expected drag coefficient;
4. An atmospheric density bias of the models compensated by a corresponding drag coefficient variation.

Due to the fact that all the analyses described in this paper included the direct solar radiation pressure with eclipses and that the Earth's albedo is not able to account for a secular along-track acceleration (Sehnal, 1981; Anselmo et al. 1983; Rubincam and Weiss, 1986; Lucchesi and Farinella, 1992), the average increase of the LARES $\langle C_D \rangle$ by 20% cannot be easily explained by other radiative drag-like perturbations acting on it, but not on Ajisai. In fact, the latter should have been much more sensitive in this respect, being its area-to-mass ratio higher by a factor 20, and the perturbation signature should have been nearly identical to that of neutral drag, in order to explain the success of the semi-major axis fit detailed in the following section. On the other hand, a sizable radiative perturbation acting on Ajisai but not on LARES would be problematic as well: in fact, a secular drag-like acceleration would have implied a further reduction of the average drag coefficient associated with neutral drag, widening the distance from LARES, while a secular acceleration in the opposite direction, partially compensating the air drag like a low thrust propulsion system, was not considered a realistic possibility. An explanation following the first hypothesis does not seem, therefore, supported by the data and results currently available.

With the goal of minimizing the eddy currents induced by the orbital motion through the Earth's magnetic field, the sub-surface of Ajisai was made of several thin layers of aluminum separated by a dielectric film (Kucharski et al., 2010). This, coupled with the silica composition of the surface, could have led to a different behavior of the satellite in the Earth's plasmasphere compared with the conductive tungsten-made LARES. A satellite acquiring an electric potential with respect to the surrounding plasma would be subjected to charged particle drag (Afonso et al., 1980; 1985), mainly due to the momentum exchange between the satellite and the interacting ions present in the atmosphere. At first sight, a significant amount of unmodeled charged particle drag on LARES, compared with Ajisai, would have resulted in a higher determined drag coefficient, as actually observed. However, the signatures of neutral and charged particle drag are quite different (Andrés de la Fuente, 2007), and solving for only the former perturbation would not have been so successful in reproducing the observed semi-major axis decay, as shown in the following section, had a substantial charged particle drag been present. On the other hand, reversing the situation, with a prevailing charged particle drag on Ajisai, would have implied a further decrease of the average drag coefficient associated with neutral drag, widening even more the discrepancy with LARES.

As pointed out at the beginning, LARES was a quite special satellite, due to the choice of a tungsten alloy to build it. However, based on the current definitions of the accommodation coefficient (Cook, 1966; Afonso et al., 1985; Pardini et al., 2010), it was impossible to explain the drag coefficient differences in terms of diverse satellite surface compositions. Even the significantly higher surface temperature of the tungsten surface – around 400 K on average – compared with a silica surface – around 300 K on average – (Brooks and Matzner, 2016) would explain, at most, a further increase of the drag coefficient by less than 2% (Mehta, 2013). Only an important deviation from the diffuse reemission of incident particles, with a sizable fraction reemitted in the opposite direction, would be able to explain the determined values. Therefore, unless some important pieces of information concerning the LARES surface properties and/or its interaction with the neutral atmosphere constituents were missing, the third hypothesis seems presently not viable.

Another possible explanation might be found in the inaccuracies of the neutral atmosphere models themselves, operating at the limits of their application range. It could not be therefore excluded that the models, in the period considered and along the orbit of LARES, underestimated the real neutral particle density, on average, by about 20%, requiring a corresponding compensation of the bias through an increase of the drag coefficient. After all, even larger density biases were inferred in the past around that altitude (Pardini and Anselmo, 2001) and comparable biases were documented as well at lower altitudes, where the thermospheric models should be more reliable, due to the distribution of the data sources used to build them (Chao et al., 1997; Pardini and Anselmo, 2001; Bowman and Moe, 2006; Pardini et al., 2010; Pardini et al., 2012).

At the LARES altitude, and for the mean exospheric temperature during the period considered, the atmosphere scale height was about 350 km (Zarrouati, 1987), so the 40 km difference between the average heights of LARES and Ajisai might seem not enough to justify a relative density bias of about 20%. However, due to their different orbital inclinations, LARES crosses high atmospheric latitudes up to 70 deg, compared with 50 deg for Ajisai, so the way the atmospheric models responded to latitudinal variations possibly played a role in addition to local vertical changes. This assumption was supported by a simple computational test, in which LARES and Ajisai were maintained at their respective height, but with switched inclination. In such a way, in order to reproduce the observed semi-major axis decay, the average drag coefficient of LARES had to be reduced by 4%, using NRLMSISE-00, while that of Ajisai had to be increased by 5%, confirming the presence of significant atmospheric density biases depending on the latitude.

Moreover, the different behavior among the models passing from the Ajisai to the LARES orbit offered further clues in support of the density bias hypothesis, because the changes observed were not uniform. Taking NRLMSISE-00 as reference, the Ajisai's $\langle C_D \rangle$ obtained with GOST-2004 and JR-71 were higher, respectively, by 4.4% and 6.6%. For LARES, the $\langle C_D \rangle$ obtained with GOST-2004 and JR-71 were higher, respectively, by 11.4% and 4.5% with respect to the NRLMSISE-00 value, increased in turn by 18.1% compared with Ajisai. The sizable relative changes among the models, up to 7% in the case of GOST-2004, are then indicative of the fact that substantial relative density biases might also occur between two orbits apparently so similar.

Based on the available data and results, it was therefore concluded that most of the observed secular semi-major axis decay of LARES and Ajisai was due to neutral atmosphere drag (or to a combination of forces acting exactly as neutral atmosphere drag). The differences found between the average drag coefficients are probably due, to a larger extent, to atmospheric density biases and, to a lesser extent, to further adjustments of the LARES accommodation coefficient accounting for its peculiar physical properties. This conclusion is fully consistent with the predictions, uncertainties and range of applicability of some of the best thermospheric density models available. Contrary to what happened in the case of LAGEOS and LAGEOS II, neutral atmosphere drag is therefore a major player among the non-gravitational perturbations acting on LARES, and its secular, long-term and short-term signatures must be modeled in detail, in order to reliably detect and characterize other comparable or smaller (depending on the component) perturbing accelerations.

**6. LARES precise orbit determination including neutral atmosphere drag**

Having deeply investigated and estimated the perturbing acceleration due to neutral atmosphere drag acting on LARES, using five different thermospheric density models, the GEODYN code was used to obtain a precise orbit determination including neutral drag, the EIGEN-GRACE02S static model for the Earth's gravity field (Reigber et al., 2005), direct solar radiation pressure with eclipses (Pavlis et al., 1998), the Earth's albedo (Rubincam et al., 1987), and the time-varying components of the geopotential (Ray, 1999; Petit and Luzum, 2010). Among the three neutral atmosphere models implemented in the available version of GEODYN, two, i.e. JR-71 and MSIS-86, were among those investigated with SATRAP and were then chosen for the analysis. Based on the discussion and the results outlined in the previous sections, MSIS-86 is much more similar to NRLMISE-90, but the overall behavior of JR-71 is anyway coherent with that of the other American models, sharing a common development history and some data sets, provided that the drag coefficient is appropriately rescaled. According to the results presented in this paper, both JR-71 and MSIS-86 can be therefore considered validated and appropriate for the LARES orbit determination with GEODYN, delivering a description of the neutral drag perturbation consistent with the predictions, uncertainties and range of applicability of all the thermospheric density models analyzed.

As in the case, mentioned at the beginning, in which atmospheric drag was not included in the dynamical model, GEODYN was again used to fit the observables over 7-day orbit arcs, from 6 April 2012 to 25 December 2015, using either JR-71 or MSIS-86. The effect on the semi-major axis residuals was dramatic (Figure 16), with a nearly complete cancellation of the unexplained decay (Lucchesi et al., 2016). This result was obtained by JR-71 with $\langle C_D \rangle$ = 3.96, and by MSIS-86 with $\langle C_D \rangle$ = 3.76, i.e. in extremely good agreement with the totally independent estimates carried out with SATRAP. In the first case the difference was, in fact, approximately 0.2%, in the second one about 1.3%, confirming again that neutral atmosphere drag alone (or a combination of forces with the same signature) can explain most of the observed semi-major axis decay of LARES.

It should also be stressed that, after modeling the neutral drag perturbation in GEODYN, a residual semi-major axis decay, corresponding to an average transverse acceleration of about $-2 \times 10^{-13}$ m/s$^2$ (i.e. $\approx$ 1/72 of neutral drag), was still detected (Lucchesi et al., 2016). Probably, the thermal thrust effects come to play a role at this level, but the value found so far is only about 50% of that reported and modeled in Nguyen and Matzner (2015) and in Brooks and Matzner (2016), so further and more detailed investigations, including the detection of the signature of the periodic terms, will be needed to characterize such smaller non-gravitational perturbations.

**7. Conclusions**

Having the experience of the two LAGEOS in mind, LARES was conceived to minimize as much as possible the impact of non-gravitational perturbations on its precise orbit determination based on laser ranging measurements. The clever design was certainly successful in achieving the highest bulk density of any known orbiting object, either natural or man-made, in the Solar System, with a correspondingly very low value of the area-to-mass ratio. However, the relatively low orbit in which LARES was placed, with a mean altitude of about 1450 km compared with about 5800-5900 km for the two LAGEOS, curtailed in part the advantages of a so low area-to-mass ratio. Therefore, while direct solar radiation pressure was smaller for LARES by a factor 2.6, the Earth's albedo radiation pressure was reduced by 50% and the effects of thermal emissions were probably significantly abated as well, concerning neutral atmosphere drag an increase of 50 times or more was expected compared with the two LAGEOS.

The main goal of this study was indeed the detailed characterization and understanding of the neutral drag perturbation on LARES. For such a task, two independent orbit propagators (GEODYN II and SATRAP), five thermospheric density models (JR-71, MSIS-86, MSISE-90, NRLMSISE-00 and GOST-2004) and two passive spherical satellites (LARES and Ajisai), just 40 km apart in altitude, were used over a 3.7 year time span, from 6 April 2012 to 25 December 2015. Based on the observed semi-major axis decay, the comprehensive analysis carried out allowed the quantitative estimation of the neutral drag acceleration components, detailing both the secular and main periodic contributions. Contrary to what had happened in the case of the two LAGEOS, it was found that neutral atmosphere drag alone (or a combination of forces having exactly the same signature) was able to explain most ($\approx 98.6\%$) of the observed semi-major axis decay of LARES and all the thermospheric density models used provided consistent results. Focusing the attention on the three most independent models, i.e. JR-71, NRLMSISE-00 and GOST-2004, the differences between the average fitted drag coefficients and their mean value were $< 6\%$, quite small compared with the acknowledged absolute density uncertainties of the models. Moreover, when the same models (JR-71 or MSIS-86) were used both in SATRAP and GEODYN, the differences between the estimated average drag coefficients were of the order of 1% or less.

The comparative analysis of the neutral atmosphere perturbation acting on Ajisai and LARES also revealed an average drag acceleration higher than expected (by approximately 20%) in the latter case. Having Ajisai an area-to-mass ratio nearly 20 times that of LARES, and being therefore much more sensitive to non-gravitational forces, like radiation pressure and thermal emission, no plausible radiative perturbation acting on LARES but not on Ajisai was found. The opposite situation, on the other hand, would have lead to still greater discrepancies. The quite different construction and surface properties of the two satellites might have also resulted in a different surface charging and electric potential with respect to the plasmasphere, causing a diverse response to charged particle drag, if present. However, even this possibility was not supported by the available results. The only reasonable explanation compatible with the results obtained, where neutral drag modeling provided an excellent fit with the appropriate drag coefficients, was that the considered thermospheric models, on average and in the adopted time interval, underestimated the real atmospheric density by about 20% along the 40 km lower LARES orbit, which probes high atmospheric latitudes up to 70 deg, compared with 50 deg for Ajisai. But even if this were the case, the neutral atmosphere models were able to explain most of the observed semi-major axis decay well within their uncertainties and applicability range.

Finally, having modeled the neutral atmosphere drag in GEODYN, a residual semi-major axis decay of LARES, corresponding to an average along-track acceleration of about $-2 \times 10^{-13}$ m/s$^2$ (i.e. $\approx 1/72$ of neutral drag), was detected as well. If linked to thermal emission effects, it was more than two orders of magnitude smaller than for the two LAGEOS, providing a striking demonstration of the LARES good design in terms of minimization of non-gravitational forces (Ciufolini et al., 2012). Again, also considering the predominant role played by neutral atmosphere drag, further and more detailed investigations, including the detection of the signature of the periodic terms, will be needed in order to characterize such smaller non-gravitational perturbations and disentangle them from other possible effects, like tiny anisotropic radiation reflection.

In conclusion, the combination of satellite design and operational orbit choice resulted for LARES in an overturning of the impact of non-gravitational perturbations compared with the two LAGEOS satellites, putting neutral atmosphere drag at center stage. Any accurate analysis of the LARES orbit and any deep investigation of small forces effects cannot then ignore a thorough scrutiny of such an important perturbation.

## Acknowledgements

This work was carried out in the framework of the LAser RAnged Satellites Experiment (LARASE) and was in part supported by the Commissione Scientifica Nazionale II (CSNII) on Astroparticle Physics Experiments of the Istituto Nazionale di Fisica Nucleare (INFN), in Italy.

The authors acknowledge the International Laser Ranging Service (ILRS) for providing high quality laser ranging data of the LARES satellite and the US Space Track Organization for making available the Two-Line Elements of LARES and Ajisai used in this study.

**Figure captions**

Figure 1. Semi-major axis decay of LARES determined by integrating the precise orbit estimation residuals over 3.7 years, obtained with GEODYN II not including in the dynamical model neither the neutral and charged atmosphere drag, nor the thermal effects.

Figure 2. LARES geodetic altitude, and ambient exospheric temperature and neutral atmosphere density, during the first 3.7 years of the mission. The exospheric temperature and the neutral atmosphere density were estimated with the NRLMSISE-00 empirical thermospheric model. The mean values were 1454 km for the geodetic altitude, 971.2 K for the exospheric temperature and $5.886 \times 10^{-16}$ kg/m$^3$ for the neutral atmosphere density.

Figure 3. Concentrations of the main neutral atomic species (He, H and O) at the LARES altitude during the first 3.7 years of the mission. They were estimated with the NRLMSISE-00 empirical thermospheric model.

Figure 4. Transverse acceleration component (S) due to neutral drag acting on LARES, compared with solar and geomagnetic activities. The eclipse periods, when part of the LARES orbit crossed the Earth's shadow, are shown in gray in the first plot.

Figure 5. Comparison of the transverse acceleration components due to neutral drag, as computed with the five models JR-71, MSIS-86, MSISE-90, NRLMSISE-00 and GOST-2004.

Figure 6. Radial (R), transverse (S) and normal (W) components of the neutral drag acceleration acting on LARES, according to the JR-71 thermospheric density model.

Figure 7. Radial (R), transverse (S) and normal (W) components of the neutral drag acceleration acting on LARES, according to the MSIS-86 thermospheric density model.

Figure 8. Radial (R), transverse (S) and normal (W) components of the neutral drag acceleration acting on LARES, according to the MSISE-90 thermospheric density model.

Figure 9. Radial (R), transverse (S) and normal (W) components of the neutral drag acceleration acting on LARES, according to the NRLMSISE-00 thermospheric density model.

Figure 10. Radial (R), transverse (S) and normal (W) components of the neutral drag acceleration acting on LARES, according to the GOST-2004 thermospheric density model.

Figure 11. Ajisai geodetic altitude, and ambient exospheric temperature and neutral atmosphere density, during the first 3.7 years of the LARES mission. The exospheric temperature and the neutral atmosphere density were estimated with the NRLMSISE-00 empirical thermospheric model. The mean values were 1494 km for the geodetic altitude, 959.5 K for the exospheric temperature and $5.213 \times 10^{-16}$ kg/m$^3$ for the neutral atmosphere density.

Figure 12. Concentrations of the main neutral atomic species (He, H and O) at the Ajisai altitude during the first 3.7 years of the LARES mission. They were estimated with the NRLMSISE-00 empirical thermospheric model.

Figure 13. Radial (R), transverse (S) and normal (W) components of the neutral drag acceleration acting on Ajisai, according to the JR-71 thermospheric density model.

Figure 14. Radial (R), transverse (S) and normal (W) components of the neutral drag acceleration acting on Ajisai, according to the NRLMSISE-00 thermospheric density model.

Figure 15. Radial (R), transverse (S) and normal (W) components of the neutral drag acceleration acting on Ajisai, according to the GOST-2004 thermospheric density model.

Figure 16. Semi-major axis cumulative residuals of LARES determined with a precise orbit estimation over 3.7 years using GEODYN II. In one data reduction the neutral atmosphere was not considered, setting to zero the drag coefficient, while in the other data reduction the neutral atmosphere was included and the drag coefficient was a solve for parameter in the differential correction estimation process. The dramatic reduction of the semi-major axis residuals confirms the overwhelming role played by neutral atmosphere drag in the case of LARES.

**Table 1**

Summary of the adjusted drag coefficients ($C_D$) able to reproduce the mean secular along-track non-gravitational acceleration observed on LARES.

| Neutral atmosphere density model | Average adjusted drag coefficient $\langle C_D \rangle$ |
|---|---|
| JR-71 | 3.95 |
| MSIS-86 | 3.71 |
| MSISE-90 | 3.73 |
| NRLMSISE-00 | 3.78 |
| GOST-2004 | 4.21 |

**Figure 1**

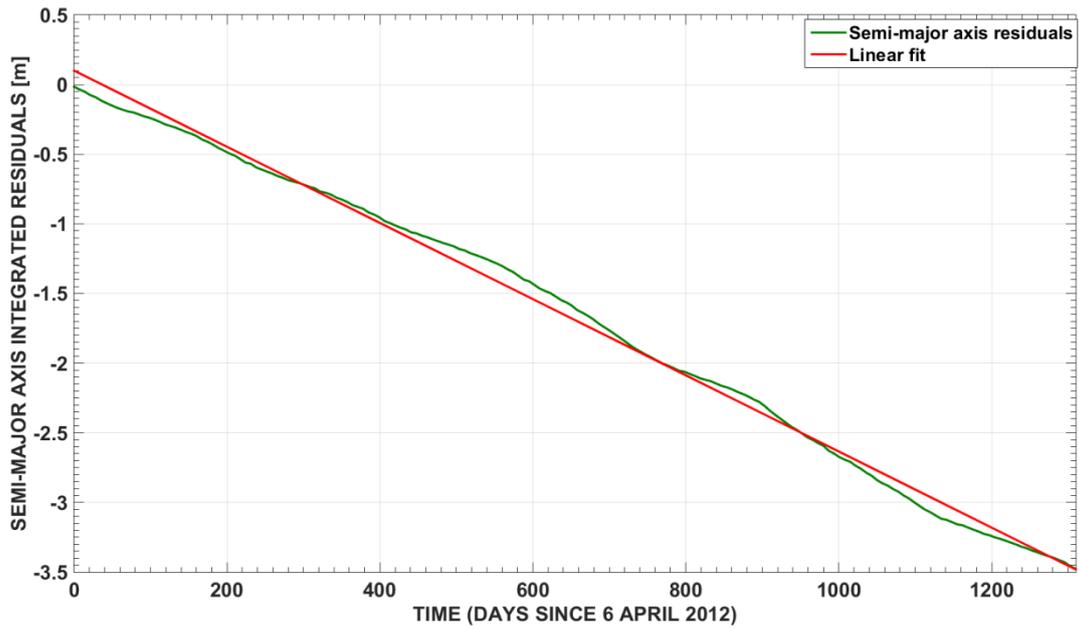

**Figure 2**

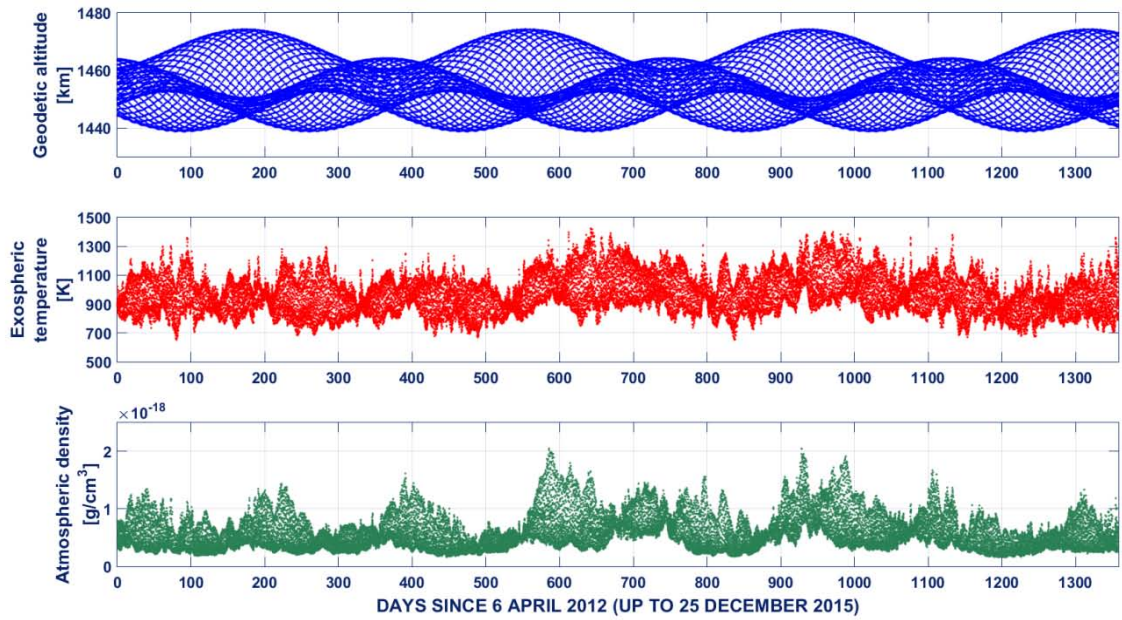

**Figure 3**

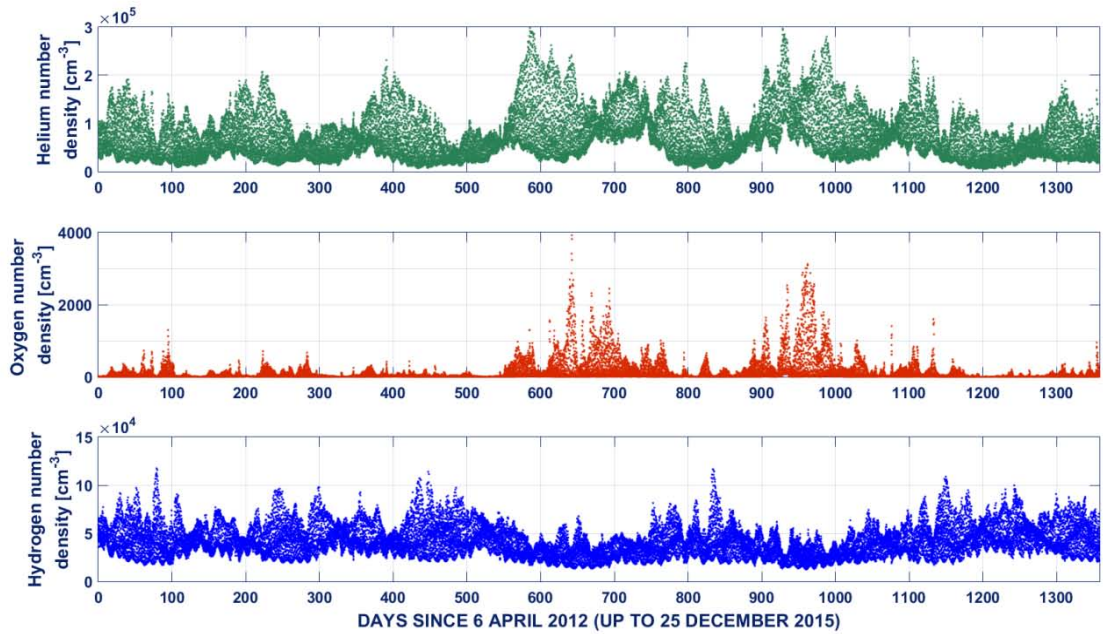

**Figure 4**

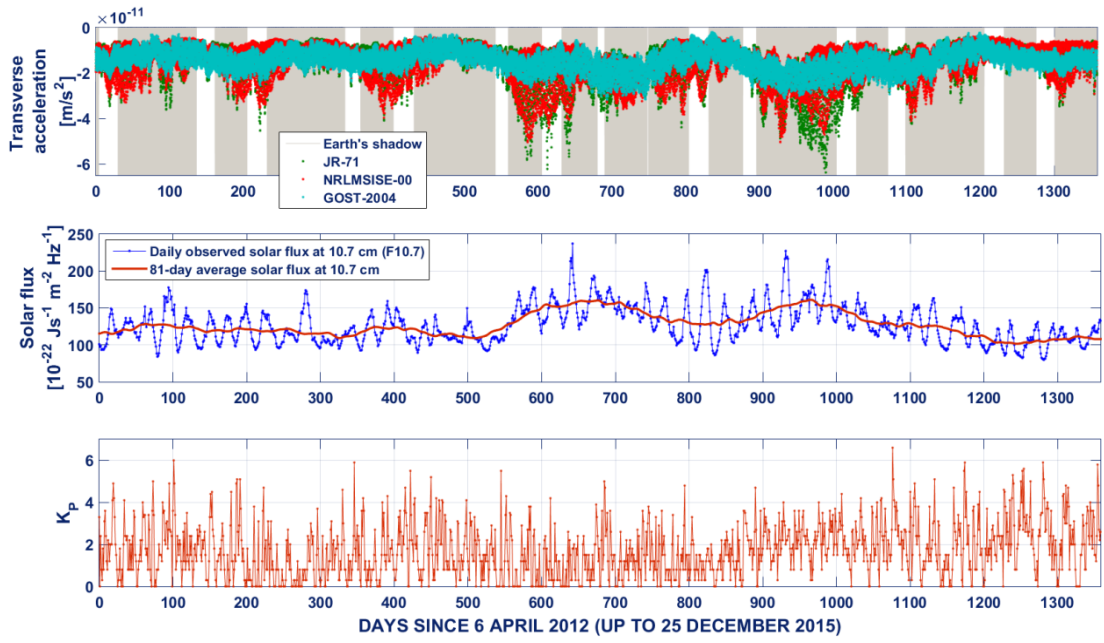

**Figure 5**

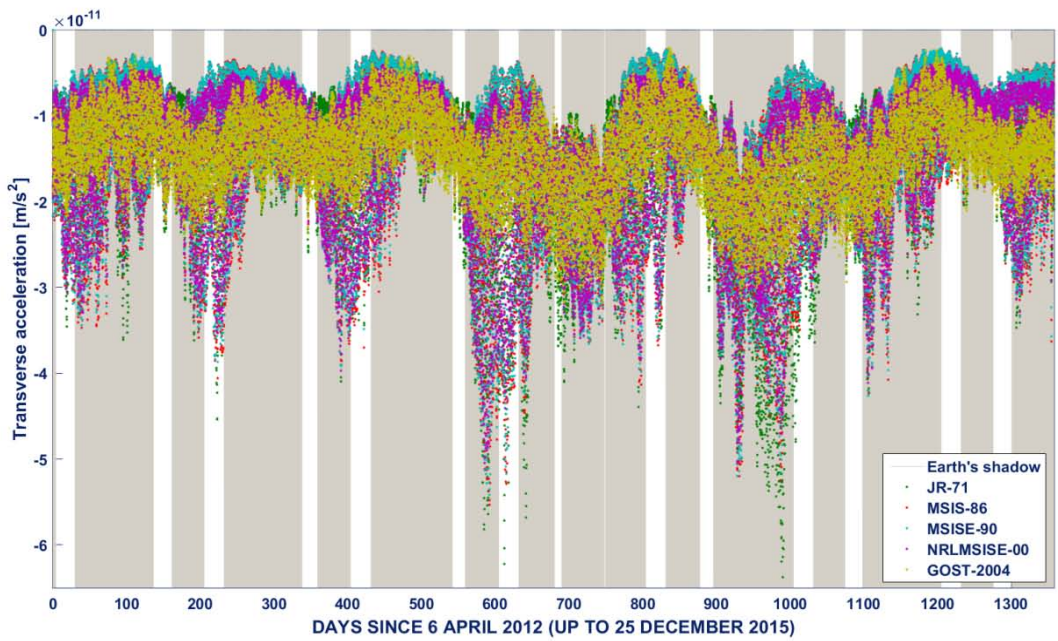

**Figure 6**

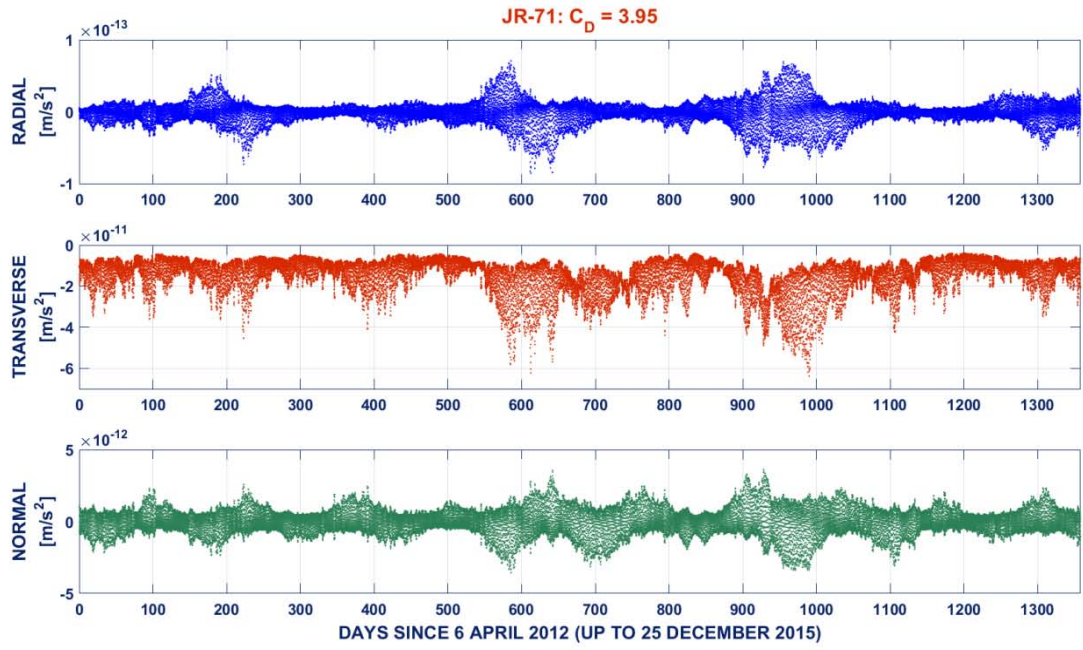

**Figure 7**

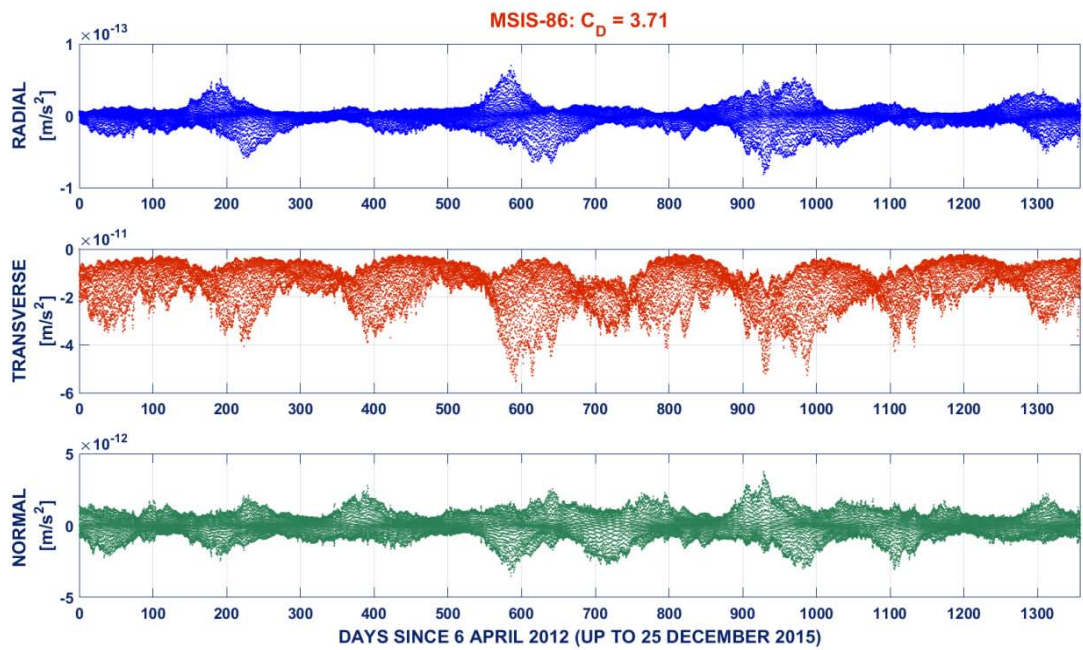

**Figure 8**

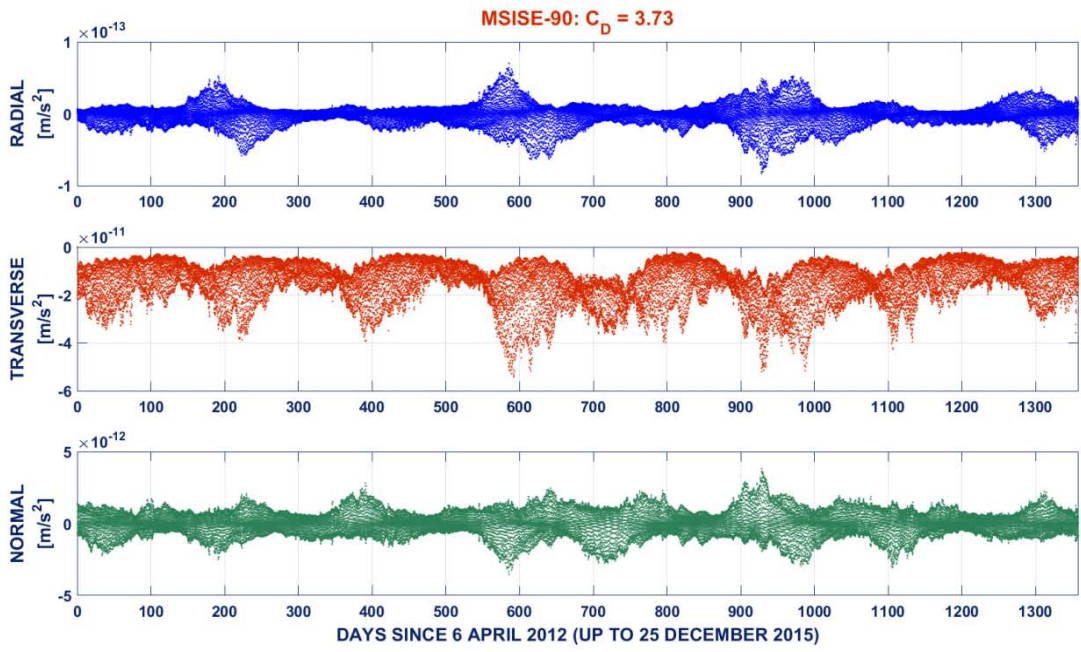

**Figure 9**

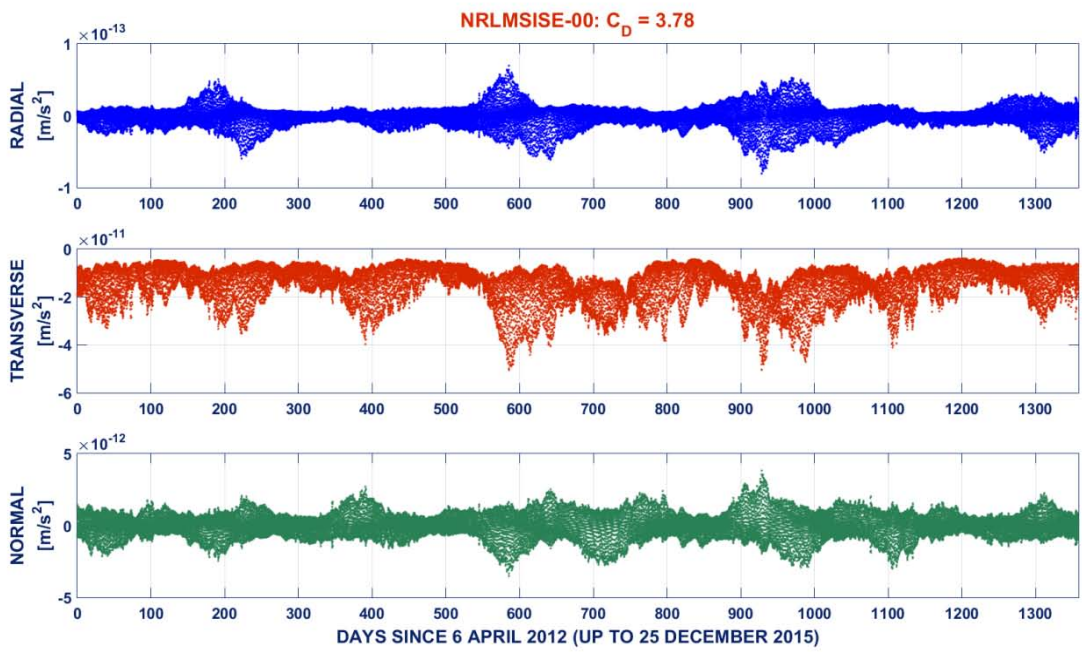

**Figure 10**

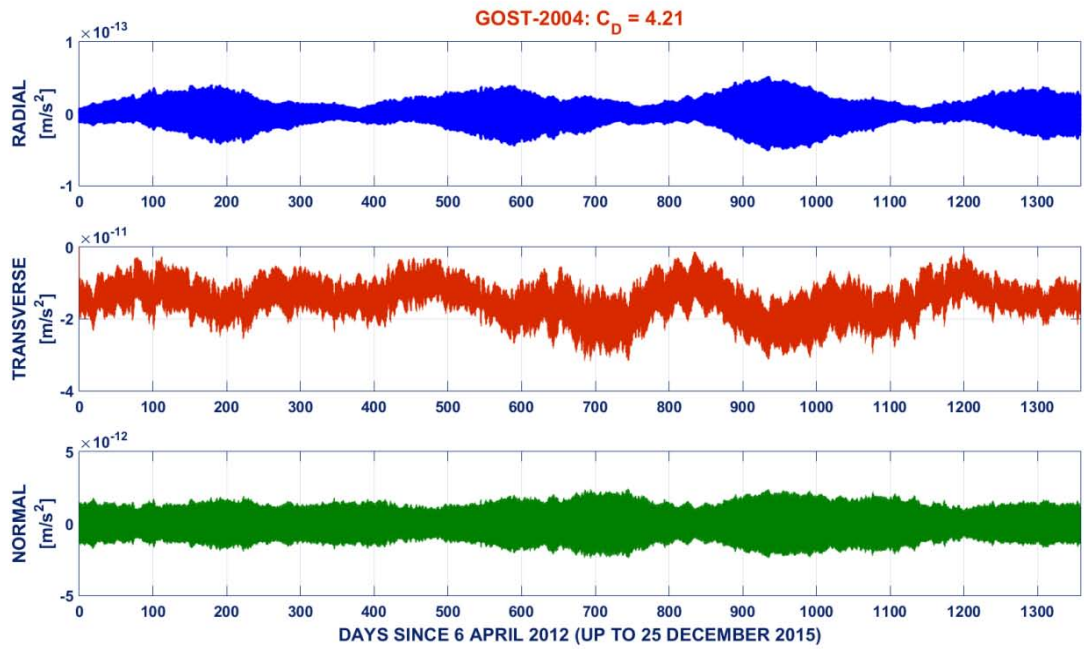

**Figure 11**

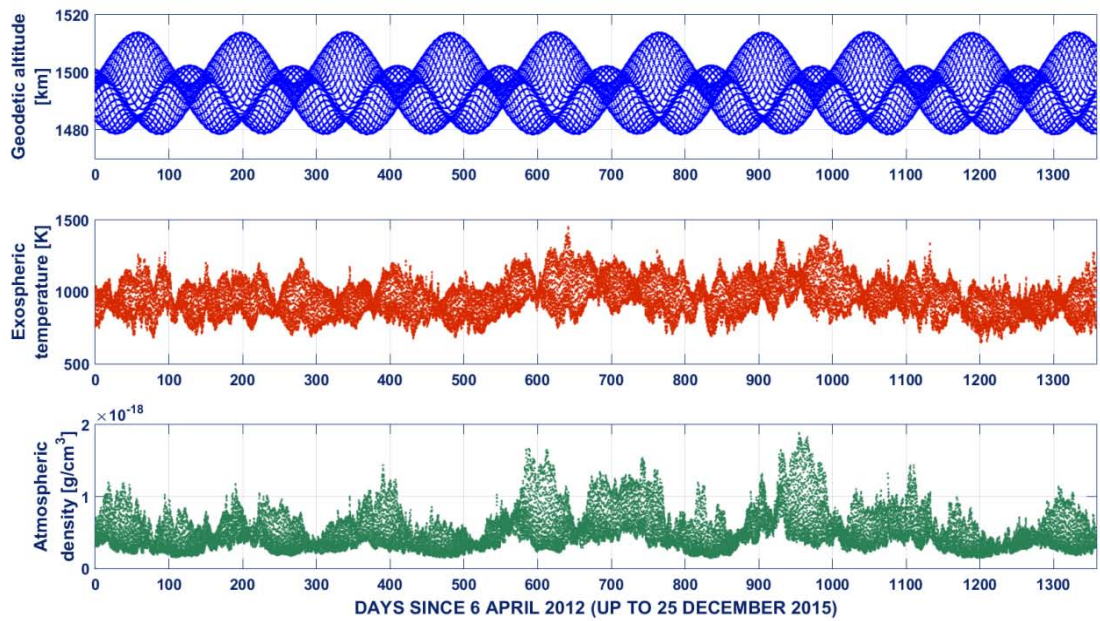

**Figure 12**

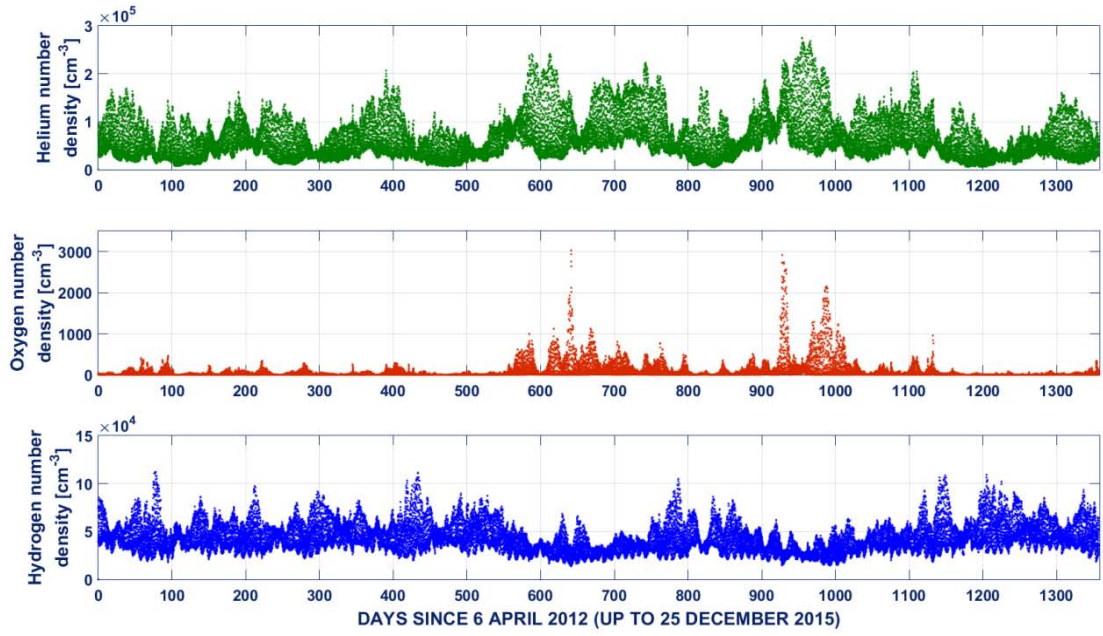

**Figure 13**

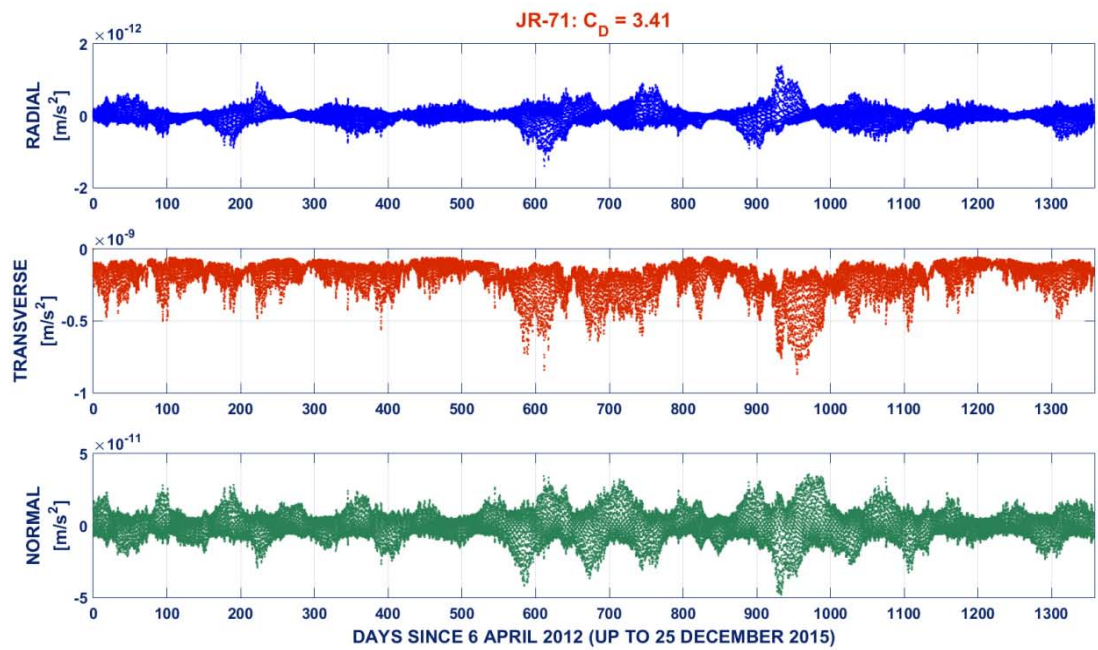

**Figure 14**

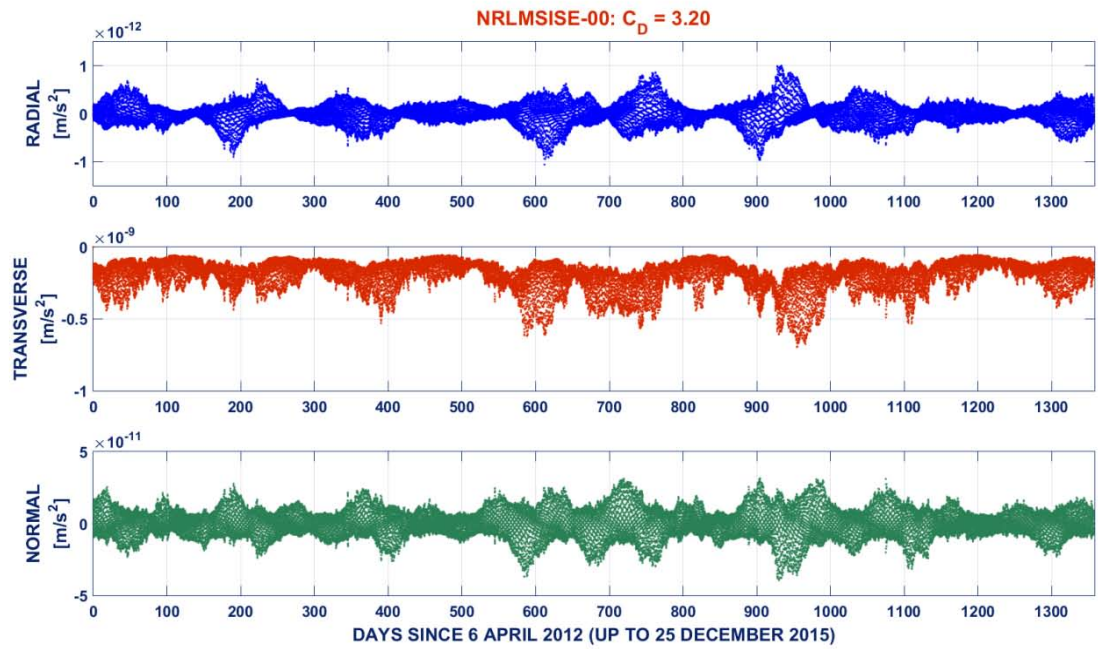

**Figure 15**

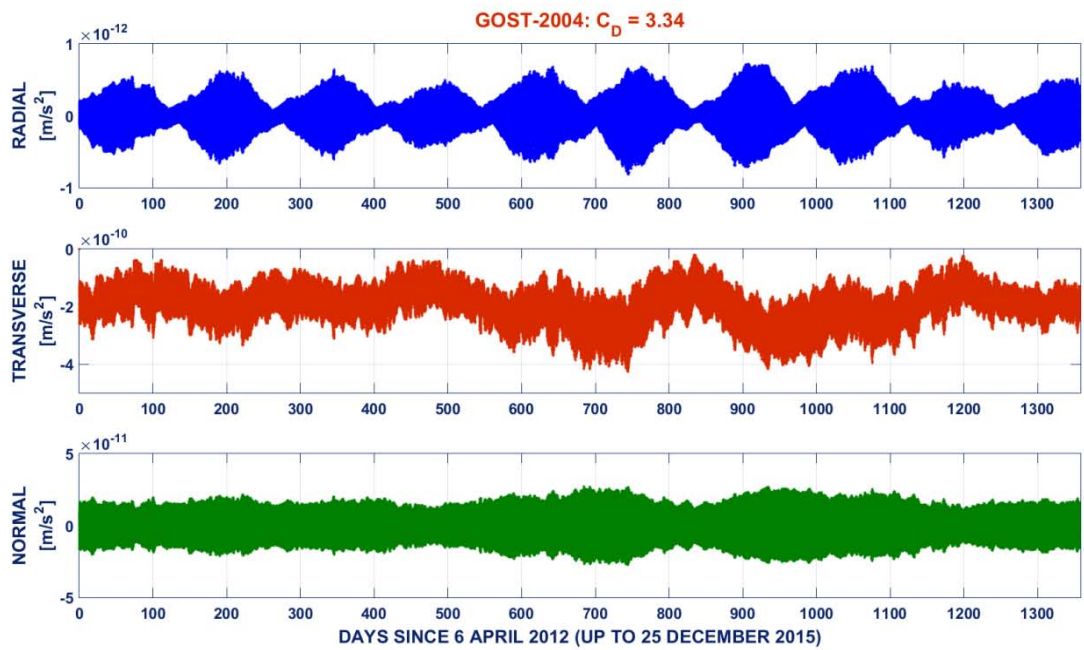

**Figure 16**

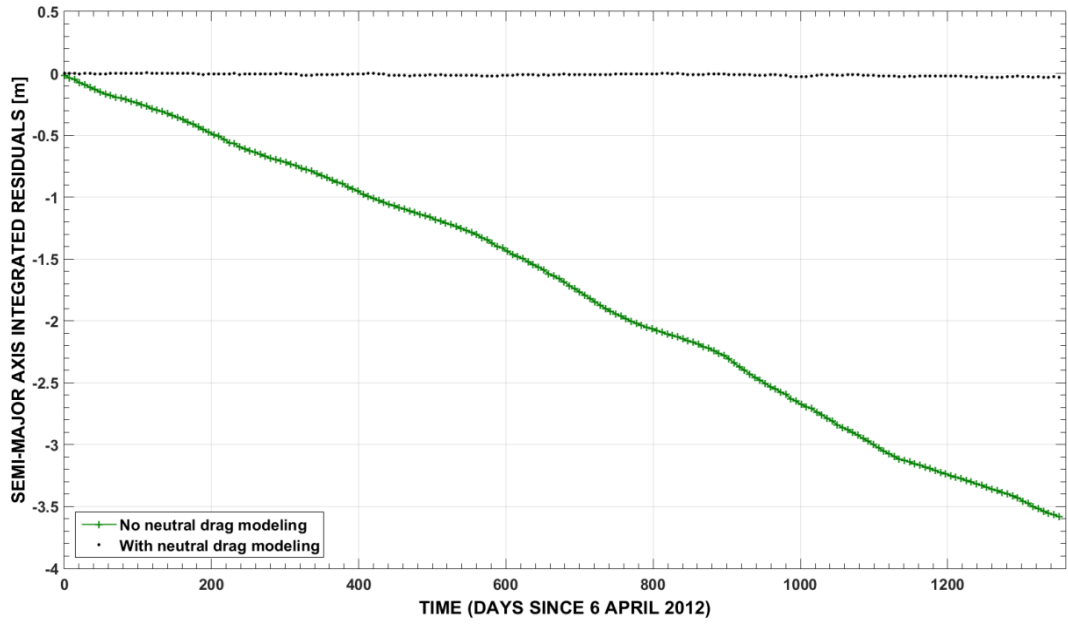